\pdfoutput=1

\documentclass[11pt]{article}

\usepackage[final]{acl}

\usepackage{times}
\usepackage{latexsym}
\usepackage{amsmath}

\usepackage[T1]{fontenc}

\usepackage[utf8]{inputenc}

\usepackage{microtype}

\usepackage{inconsolata}

\usepackage{graphicx}
\usepackage{hyperref}

\usepackage{listings}
\usepackage{xcolor}

\title{Idiosyncratic Versus Normative Modeling \\of Atypical Speech
Recognition: Dysarthric Case Studies}

\author{Vishnu Raja\thanks{Equal contribution}\\Stony Brook University \And Adithya V Ganesan\footnotemark[1]\\Stony Brook University \And Anand Syamkumar\\Stony Brook University \AND Ritwik Banerjee\\Stony Brook University \And H. Andrew Schwartz\\Vanderbilt University\\Stony Brook University \AND \tt{\{visraja, avirinchipur, asyamkumar, rbanerjee, has\}@cs.stonybrook.edu}}

\begin{document}
\maketitle
\begin{abstract}

State-of-the-art automatic speech recognition (ASR) models like Whisper, perform poorly on atypical speech, such as that produced by individuals with dysarthria. 
Past works for atypical speech have mostly investigated fully personalized (or idiosyncratic) models, but modeling strategies that can both generalize and handle idiosyncracy could be more effective for capturing atypical speech. 
To investigate this, we compare four strategies: (a) \textit{normative} models trained on typical speech (no personalization), 
(b) \textit{idiosyncratic} models completely personalized to individuals, 
(c) \textit{dysarthric-normative} models trained on other dysarthric speakers,
and 
(d) \textit{dysarthric-idiosyncratic} models which combine strategies by first modeling normative patterns before adapting to individual speech.
In this case study, we find the dysarthric-idiosyncratic model performs better than idiosyncratic approach while requiring less than half as much personalized data (36.43 WER with 128 train size vs 36.99 with 256). 
Further, we found that tuning the speech encoder alone (as opposed to the LM decoder) yielded the best results reducing word error rate from 71\% to 32\% on average.
Our findings highlight the value of leveraging both normative (cross-speaker) and idiosyncratic (speaker-specific) patterns to improve ASR for underrepresented speech populations.\footnote{\href{https://github.com/VishnuRaja98/Dysarthric-Speech-Transcription}{Github: VishnuRaja98/Dysarthric-Speech-Transcription}}
\end{abstract}

\section{Introduction}

\begin{figure}[t]
\begin{center}
\includegraphics[width=0.7\linewidth]{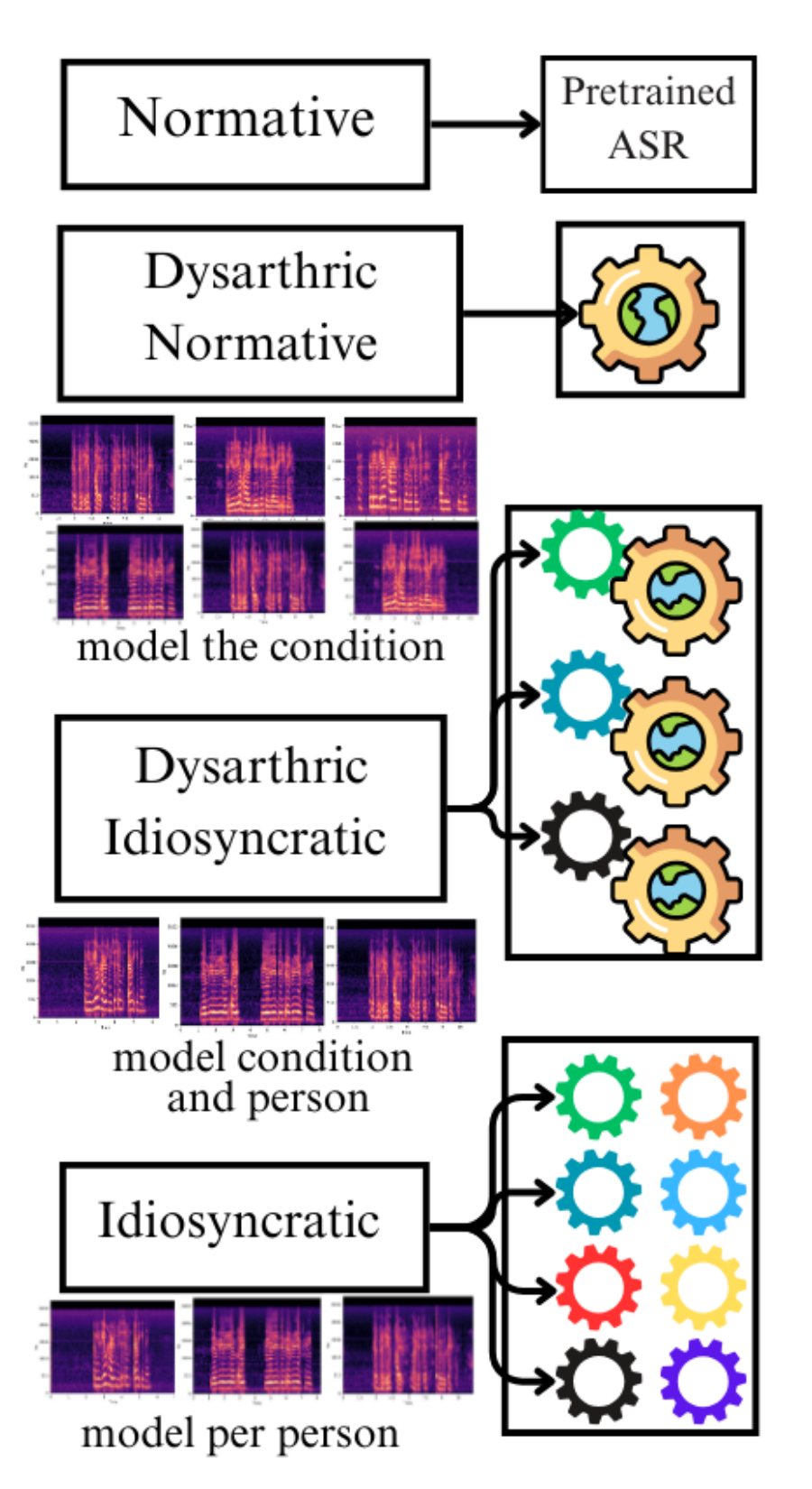}
\end{center}
\caption{Four types of models.
Normative model is Whisper-small and we do one-shot predictions on it. 
Idiosyncratic models create a model for each user.
Dysarthric Normative model, creates one model for a user while excluding it and using evry other user for cross validation.
Dysarthric Idiosyncratic model uses a users normative model and personalizes it.
}
\label{fig:methodology}
\end{figure}

ASR models are predominantly trained on normative populations, failing to generalize on individuals with atypical speech, such as dysarthria. 
Past works addressing this predominantly take an idiosyncratic modeling approach by training (or fine-tuning) separate models, one specific to each individual~\citep{shor2019personalizing, green2021automatic}. 
On top of requiring vast amounts of data from the individual, such idiosyncratic models might fail to capture the speaker's changing speech characteristics over time, eventually causing the model to generalize poorly~\citep{tomanek2023analysis}.
Alternatively, learning from the cross-section of dysarthric individuals could allow the model to adapt to individual's changing patterns within practical sample sizes.

In this work, we compare models with differing degrees of idiosyncratic (personalized) versus normative (the same for all) models. 
Specifically, we compare the performance of four strategies:
(a) normative models trained on typical speech
(b) idiosyncratic models, i.e., normative models tuned to individuals,
(c) dysarthric normative models, i.e., normative models tuned to dysarthric population,
(d) dysarthric idiosyncratic models, i.e., dysarthric normative models tuned to individuals.
The difference in performances between (c) and (a) informs the contributions of learned speech characteristics of dysarthria, whereas the difference between (d) and (c), and (d) and (b) shows the contribution of speaker-specific characteristics.  

Dysarthria is a motor speech disorder caused by damage to the nervous system, making it difficult for individuals to control and coordinate the muscles involved in speech. 
People with dysarthria often have trouble clearly pronouncing words, resulting in production of unclear speech from slurred, stuttered or arrhythmic patterns.   
This difference in the acoustic signal between dysarthric individuals and the normative population, causes normative models to fail.
However, dysarthria does not affect a person's ability to think or understand language; rather, it affects their ability to physically produce speech like normative population due to muscle weakness or lack of coordination.
With the typical size of dysarthria speech datasets ranging in 10s of people\footnote{The UASpeech corpus~\citep{kim2008uaspeech} comprises 19 speakers with dysarthria, while the TORGO dataset~\citep{rudzicz2012torgo} used in this work comprises 15 speakers -- 8 with dysarthria, and 7 for control. 
Others corpora, including those for non-English speakers, offer similar sizes: 31 dysarthric speakers in the Italian-language EasyCall dataset~\citep{turrisi2021easycall}, 44 in the Chinese-language CDSD corpus~\citep{cdsd2023}, and 30 in the Tamil SSNCE corpus~\citep{tamil2016}.}, it is more viable to leverage transfer learning of normative models rather than adopting an extremely challenging approach of training a model from scratch.
With supporting evidence~\citep{goldstein2025unified, tuckute2024driving, aw2024instructiontuning} for shared activation regions between human brain and deeper layers of speech \& language models, normative models can be adapted to capture the differences in surface form speech patterns to map back to the same activation regions of language, thus avoiding the need to train a model from scratch.

The scaling trends have directed the models into 100s of millions of parameters with growing number of layers, hidden dimensions, and the normative data it was trained with~\citep{kaplan2020scaling, hoffmann-etal-2022}, whilst maintaining the poor performances on underrepresented population. 
Adapting these large scale models using transfer learning, especially with characteristic drift in the speech signals from normative population would require parameter efficient approaches~\citep{hu2022lora, v-ganesan-etal-2021-empirical}.
To this, we compare two parameter efficient strategies of training ASR models against standard full fine-tuning, namely, only tuning the speech encoder and only tuning the language decoder to quantify its effect on performance. 

Our main contributions include: 
(1) Systematic comparison between different ways to improve dysarthric speech recognition model to quantify the contributions of dysarthric speech characteristics and person-specific speech characteristics.
(2) a parameter efficient approach to fine-tune ASR models to achieve the best performance
(3) Analysis of the different models' WER against individuals' severity scores of motor functions.
We found that:
(a) 30.5\% of performance improved from learning dysarthric speech characteristics, and 23.57\% improved from learning speaker specific characteristics.
(b) training the speech encoder part of the ASR models led to consistent improvements over full fine-tuning or language encoder alone for all adaptation strategies.
(c) the improvements in the ASR model for dysarthric speech corresponded to decreased correlation with motor control severity scores of the individuals.  
These findings on whisper-small generalized to whisper-medium, showing that the results hold even with scaling up the model size.

\section{Related Work}
\label{sec:related-work}
In recent years, ASR has achieved remarkable progress in the detection of atypical speech patterns, primarily through alignment-based data augmentations~\citep{xiong2019phonetic}, contrastive learning~\citep{wu2021sequential, yang2025feature}, and self-supervised learning with augmentations to both data and deep neural architectures~\citep{hu2024self, takashima2024self, takashima2024dysarthric}.
Self-supervised learning was significantly aided by the introduction of wav2vec~\citep{baevski2020wav2vec}, leading to demonstrable improvements in atypical speech recognition and severity assessments~\citep{javanmardi2023wav, javanmardi2024exploring, nguyen2024exploring}.
Despite this success, some reports indicate that supervised learning maintains superior performance in pathological speech recognition (for example, \citet{violeta2022investigating} and \citet{baskar2022speaker}).

As such, a sizable body of work has explored the use of large-scale ASR models trained on typical speech and subsequently fine-tuned on small atypical speech corpora~\citep{shor2019personalizing, doshi2021extending, green2021automatic}. 
Further, to overcome concerns of data paucity, efficient adaptations have demonstrably improved atypical speech detection through the use of residual adapters~\citep{tomanek2021residual} and transfer learning with small amounts of cohort data~\citep{tomanek2021on}, or a fusion of cohort and individual data~\citep{qi2023interspeech}.

Recent improvements have largely followed this two-stage methodology: employing an ASR model pretrained on general speech for fine-tuning with cohort-level data, and then individual personalization~\citep{takashima2020two, muller-eberstein2024}.
As this approach is theoretically grounded in knowledge transfer principles, it echoes earlier work leveraging transfer-learning~\citep{vachhani2017deep, takashima2020dysarthric} as well as more recent explorations that use meta-learning~\citep{wang2021improved, hu2022neural} and few-shot learning~\citep{hermann2023few} to demonstrate that even limited idiosyncratic (\textit{i.e.}, speaker-specific) data can improve speech recognition for dysarthric speakers.
In a similar vein, the recent work by \citep{hsieh2024multimodal} and \citep{qi2025model} explore the utility of curriculum learning by combining phonological features with model representations and traditional acoustic features. 
For a comprehensive review of studies on dysarthric speech and ASR systems, we point the reader to the recent survey by \citep{bhat2025speech}.

Despite these advances in adapting to dysarthric speech, the critical interplay between speech encoder specialization and efficient use of data remains underexplored.
Existing frameworks often overlook systematic evaluation of modular adaptations, particularly the counterproductive effects of language model decoder tuning observed in our work.
Our findings not only challenge prevailing adaptation strategies but also empirically establish encoder-focused tuning and hybrid cohort-idiosyncratic learning as superior paradigms, advancing both performance and practicality in low-resource clinical settings.

\section{Dataset}\label{sec:dataset}

In this work, we use the TORGO database, a collection of acoustic and articulatory speech data from individuals with dysarthria caused by either cerebral palsy or amyotrophic lateral sclerosis~\citep{rudzicz2012torgo}.
All participants read English text from a screen displaying prompts, which included short words, sentences, images (described by the participants), and non-words resembling speech sounds.
There were 8 participants in total, labeled as F01, F03, F04, M01, M02, M03, M05, and M05. 
Note that F02 is not present in any of the experiments.
The dataset includes speech from eight dysarthric speakers, whose motor functions were assessed using the \textbf{Frenchay Dysarthria Assessment (FDA) }~\citep{enderby2011}.
The FDA evaluates 28 perceptual dimensions of speech, categorized into reflex, respiration, lips, jaw, soft palate, laryngeal function, tongue function, and intelligibility~\ref{tab:fda-scores}.

This dataset is, to the best of our knowledge, uniquely suitable for studying naturalistic dysarthric speech despite its smaller scale. 
Unlike the other popular choice for dysarthric speech, such as the UASpeech dataset~\citep{kim2008uaspeech}, which relies heavily on isolated word prompts, TORGO includes diverse language utterance types including spontaneous natural language elicited from images (among also including, e.g., short words and restricted sentences). 
Thus, this dataset enables us to model ASR performance in scenarios closer to real-world communication e.g., conversational fragments. 

\begin{table*}[ht]
\centering
\begin{tabular}{lcccccccc}
\hline
\bf Category & \bf F01 & \bf F03 & \bf F04 & \bf M01 & \bf M02 & \bf M03 & \bf M04 & \bf M05 \\
\hline
Reflex      & 8.0 & 6.7 & 6.7 & 8.0 & 8.0 & 7.7 & 7.0 & 7.3 \\
Resp.       & 5.0 & 8.0 & 8.0 & 3.0 & 3.0 & 7.5 & 3.0 & 1.5 \\
Lips        & 5.6 & 8.0 & 8.0 & 5.0 & 5.0 & 7.8 & 3.2 & 3.6 \\
Jaw         & 5.5 & 8.0 & 8.0 & 8.0 & 8.0 & 8.0 & 5.0 & 8.0 \\
Palate      & 5.3 & 8.0 & 8.0 & 6.7 & 6.7 & 8.0 & 7.3 & 7.3 \\
Laryngeal   & 3.0 & 8.0 & 8.0 & 2.5 & 2.5 & 7.0 & 2.3 & 4.5 \\
Tongue      & 2.3 & 6.7 & 6.7 & 2.3 & 2.3 & 7.7 & 3.3 & 2.2 \\
Intel.      & 2.3 & 8.0 & 8.0 & 2.3 & 2.3 & 8.0 & 1.7 & 5.3 \\
\hline
\bf SUM     & 37.1 & 61.3 & 61.3 & 37.8 & 37.8 & 61.6 & 32.8 & 39.8 \\
\hline
\end{tabular}
\caption{Frenchay Dysarthria Assessment (FDA) for all users across speech-related categories, 
an 8 point scale with 8 corresponding to normal speech and 1 corresponding to severely affected speech.}
\label{tab:fda-scores}
\end{table*}

For this study, non-textual prompts—images and non-words—were removed during the data-cleaning process.
The final dataset consists of approximately 132 minutes of audio data, encompassing all speakers.
There are a total of 482 unique prompts, and each speaker’s speech was split into three parts for training, development, and testing to ensure that the training data from one user does not contaminate the test data of another.

To prevent data leakage, we ensured that the test prompts of one user were not seen during training, even though the model was trained on data from all users.
We randomly split the prompts in an 80-20 ratio, reserving 385 for training and 97 prompts for testing.
The train-validation split was then performed over the train audio-text pairs using the same 80-20 ratio.

\section{Experiments}\label{sec:experiments}

\subsection{Methodology}\label{sec:methodology}


We evaluated three adaptation approaches to the baseline (Off-the-shelf pre-trained Whisper) as shown in Figure~\ref{fig:methodology}:
\textbf{(1) Idiosyncratic Model:} Fine-tuned ASR models on individual users' data. 
Each model was tested on both - its target speaker (within-user evaluation) and other speakers (cross-user evaluation) to assess generalization capabilities.
\textbf{(2) Dysarthric-Normative Models:} Developed through leave-one-out cross-validation (LOOCV), where we trained on data from all but one speaker and selected each of the remaining speakers for cross-validation.
\textbf{(3) Iterative Combined Integration (ICI) Models:} Further adapted the Dysarthric-Normative models to individual speakers using limited target-user data.

To identify critical components for dysarthric speech recognition, we compared three tuning configurations:
(1) Full Model finetuning to update all parameters
(2) Encoder-Only finetuning to modify only the speech feature extractor (or preserves language processing) and
(3) Decoder-Only finetuning to adapt only the language model component (or preserves acoustic patterns)

We also measured the effect of data by progressively increasing training data starting with 16 prompts, doubling until 128 for each user.
The incremental data experiment was performed on both the base normative and the pre-adapted dysarthric normative models. 
This tests real-world feasibility given the practical challenges of collecting large dysarthric speech samples.

\subsection{Model Training Parameters}

We utilized Whisper small model from OpenAI~\citep{radford2022robustspeechrecognitionlargescale}, a transformer-based encoder-decoder model optimized for speech recognition tasks. 
Training was conducted on a combination of NVIDIA T4 and RTX A6000 GPUs, with a total compute time of 160 GPU hours. 
The model employs a micro-batch size of 2 samples per GPU with gradient accumulation steps of 4, resulting in an effective batch size of 8.

Optimization was performed using the AdamW optimizer with a learning rate of 1e-5 and mixed precision training (bfloat16) for efficiency. 
The training protocol consisted of 7 epochs with a 10\% warm-up ratio. 
Model selection was based on validation Word Error Rate (WER), with generated sequences limited to 50 tokens.

To address potential overfitting due to limited dysarthric data, we conducted experiments with various regularization techniques. 
L2 weight decay was tested with values of 0.1, 0.01, and 0.001 on the development set.
Additionally, attention dropout rates of 0.05 and 0.01 were evaluated. 
These regularization parameters were systematically varied to balance model capacity with the constraints of limited training data.

The model was trained for seq2seq generation task using the Hugging Face~\citep{wolf-etal-2020-transformers} library, with key configurations including per-device train batch size, gradient accumulation steps, learning rate, number of training epochs, mixed precision settings, and the metric for model selection (WER). 
This approach allowed for efficient utilization of computational resources while exploring the impact of different regularization strategies on model performance.

\subsection{Evaluation Procedure}

We evaluated model performance using Word Error Rate (WER), calculated as $\text{WER} = \frac{\text{S} + \text{I} + \text{D}}{\text{N}} \times 100\%$, where S, I, and D represent substitutions, insertions, and deletions, respectively, and N is the total number of words in the reference transcript. 
Text normalization was applied using the \texttt{jiwer} library, including case normalization, contraction expansion, punctuation removal, and whitespace standardization. 
Note that the values of WER can go above 100 depending on how many of S, I and D were made compared to N.

\section{Results}\label{sec:results}

\subsection{Idiosyncratic Models}

\begin{table*}[t]
\centering
\resizebox{\textwidth}{!}{
\begin{tabular}{@{}lccccccccr@{}}
\hline
\multicolumn{1}{c}{\bf Trained on} & \multicolumn{8}{c}{\bf Tested on →} & \\ 
\cline{2-9}
& \bf F01 & \bf F03 & \bf F04 & \bf M01 & \bf M02 & \bf M03 & \bf M04 & \bf M05 & \bf Row Avg \\ 
\hline
F01 & \textbf{47.22} & 38.35 & 15.32 & 76.47 & 68.33 & 12.74 & 88.82 & 85.71 & 54.12 \\
F03 & 77.78 & \textbf{30.82} & 13.06 & 75.40 & 62.78 & 10.38 & 88.82 & 78.57 & 54.70 \\
F04 & 69.44 & 37.28 & \textbf{9.91} & 71.66 & 63.33 & 8.96 & 81.58 & 75.00 & 52.14 \\
M01 & 63.89 & 40.14 & 13.06 & \textbf{47.59} & 68.33 & 12.74 & 84.87 & 71.43 & 50.25 \\
M02 & 69.44 & 39.07 & 12.16 & 70.05 & \textbf{38.89} & 11.32 & 91.45 & 75.00 & 50.92 \\
M03 & 69.44 & 42.29 & 10.36 & 78.07 & 65.00 & \textbf{8.02} & 90.79 & 71.43 & 54.42 \\
M04 & 58.33 & 40.14 & 14.86 & 66.31 & 62.78 & 16.51 & \textbf{55.92} & 78.57 & 49.17 \\
M05 & 83.33 & 40.50 & 11.26 & 65.24 & 59.44 & 12.26 & 75.00 & \textbf{57.14} & 50.52 \\
\hline
Col Avg & 67.36 & 38.57 & 12.50 & 68.85 & 61.11 & 11.61 & 82.16 & 74.11 &  \\
\hline
\end{tabular}}
\caption{Cross-User Generalization Results (WER \%) with full model Finetuning (incl. encoder and decoder). Lower is better. An idiosyncratic model is trained over each user and that model is tested for all users.}\label{cross-user-table-full-finetune}
\end{table*}

\begin{table*}[t]
\centering
\resizebox{\textwidth}{!}{
\begin{tabular}{@{}lccccccccr@{}}
\hline
\multicolumn{1}{c}{\bf Trained on} & \multicolumn{8}{c}{\bf Tested on →} & \\ 
\cline{2-9}
& \bf F01 & \bf F03 & \bf F04 & \bf M01 & \bf M02 & \bf M03 & \bf M04 & \bf M05 & \bf Row Avg \\ 
\hline
F01 & \textbf{41.67} & 39.78 & 11.26 & 84.49 & 68.33 & \textbf{6.13} & 135.53 & 107.14 & 61.79 \\
F03 & 63.89 & \textbf{30.47} & 11.71 & 80.75 & 62.78 & 9.91 & 115.79 & 75.00 & 56.29 \\
F04 & 72.22 & 39.78 & \textbf{9.01} & 73.80 & 63.33 & 9.43 & 87.50 & 89.29 & 55.55 \\
M01 & 58.33 & 42.29 & 13.96 & \textbf{43.32} & 68.33 & 6.60 & 83.55 & 89.29 & 50.71 \\
M02 & 69.44 & 37.99 & 15.77 & 75.40 & \textbf{37.78} & 9.43 & 86.18 & 71.43 & 50.43 \\
M03 & 83.33 & 45.88 & 14.41 & 81.82 & 65.00 & 6.60 & 100.00 & 82.14 & 59.90 \\
M04 & 72.22 & 42.29 & 11.71 & 64.71 & 62.78 & 13.21 & \textbf{59.21} & \textbf{64.29} & 48.80 \\
M05 & 83.33 & 39.43 & 10.36 & 71.66 & 59.44 & \textbf{6.13} & 84.87 & \textbf{64.29} & 52.44 \\
\hline
Col Avg & 68.05 & 39.74 & 12.27 & 71.99 & 60.97 & 8.43 & 94.08 & 80.36 &  \\
\hline
\end{tabular}}
\caption{Cross-User Generalization Results (WER \%) with Encoder only Finetuning. Lower is better. An idiosyncratic model is trained over each user and that model is tested for all users.}\label{cross-user-table-encoder-finetune}
\end{table*}

In our first experiment, we trained idiosyncratic models by fine-tuning a base normative model. 
The results for full fine-tuning are presented in \autoref{cross-user-table-full-finetune}, while \autoref{cross-user-table-encoder-finetune} shows the results for encoder only fine-tuning.

A key observation from these results is that the best performance in each column is consistently found along the diagonal, where the test and train data come from the same user. 
Additionally the diagonal values are always equal to or better than those of the base normative model (See \autoref{model-performance-table}).

To assess how well the models generalize across users, we analyze the row averages. 
The mean of these row averages is 54.49 for models where only Speech is tuned and 52.03 for Speech+LM tuned models, with standard deviations of 4.70 and 2.14, respectively. 
These results suggest that, for this user set, fully fine-tuned models achieve slightly better one-to-one cross-user generalization compared to encoder only finetuned models.

However, one notable exception is observed when encoder-finetuned models are tested on M03, where performance does not follow the expected trend~(\autoref{cross-user-table-encoder-finetune}). 
This could be attributed to M03’s clearer speech~(\autoref{tab:fda-scores}), making personalized adaptation less necessary.
Interestingly, the models trained on F01 and M05, who have more severe dysarthria based on their FDA scores, generalize better than models trained on M03. 
This raises the possibility that severe dysarthric speech patterns might provide more distinctive cues for adaptation compared to milder dysarthria. 
Investigating whether models trained on highly dysarthric speech can better recognize mild dysarthria could be a valuable direction for future research.



\subsection{Dysarthric Normative Models}

\begin{table*}[t]
\centering
\begin{tabular}{@{}lccc@{}}
\hline
\bf Model & \multicolumn{1}{c}{\bf Speech \& LM} & \multicolumn{1}{c}{\bf Speech Only} & \multicolumn{1}{c}{\bf LM Only} \\ 
\hline
Normative & & \textbf{70.94} & \\ 
Idiosyncratic & 36.58 & \textbf{36.54} & 54.23 \\
Dysarthric Normative & 58.19 & \textbf{49.30} & 64.44 \\ 
Dysarthric Idiosyncratic & 46.96 & \textbf{32.58} & 46.82 \\ 
\hline
\end{tabular}
\caption{Average WER\% over each models when tuning different parts of Whisper-small. 
For Dysarthric Normative, all normative models of all users are averaged and for Dysarthric Idiosyncratic, only the best of all models is chosen.}\label{architecture-performance-table}
\end{table*}

\begin{table*}[t]
\centering
\begin{tabular}{@{}lccc@{}}
\hline
\bf Model & \multicolumn{1}{c}{\bf Speech \& LM} & \multicolumn{1}{c}{\bf Speech Only} & \multicolumn{1}{c}{\bf LM Only} \\ 
\hline
Normative & & \textbf{61.38} & \\ 
Idiosyncratic & 33.93 & \textbf{31.14} & 50.25 \\
Dysarthric Normative & 53.19 & \textbf{45.51} & 60.24 \\ 
Dysarthric Idiosyncratic & 39.96 & \textbf{28.40} & 44.49 \\ 
\hline
\end{tabular}
\caption{Average WER\% over each models when tuning different parts of Whisper-medium. 
For Dysarthric Normative, all normative models of all users are averaged and for Dysarthric Idiosyncratic, only the best of all models is chosen.}\label{architecture-performance-table-medium}
\end{table*}

To further improve performance, we developed 56 dysarthric normative models using a leave-one-out approach. 
For each model, one user was excluded from training, and an additional user was excluded for validation.
For each excluded user, every other user was used for cross-validation once.
Each normative model was trained using the remaining six users' training data, and WER was calculated using the omitted user's test data.

\autoref{model-performance-table} presents the WER scores for these models, demonstrating significant improvement over the base normative model (WER 70.94; \autoref{model-performance-table}). 
On average, the dysarthric normative models reduced WER to 49.30, showing improvements across all users except F04 and M03 (\autoref{model-performance-table}). 
Notably, for F04, the performance remained unchanged, while for M03, the dysarthric normative model performed slightly worse than the base normative model.

According to \autoref{architecture-performance-table}, the best results were obtained when only the speech component was fine-tuned, rather than incorporating the language model (LM). 
The results shown in \autoref{model-performance-table} reflect this optimal configuration.

These findings indicate that learning dysarthric speech patterns from multiple users, while excluding the target user, is an effective strategy. 
The results suggest that training on speech alone—without LM adaptation—provides the most robust dysarthric normative models.





\subsection{Dysarthric Idiosyncratic models}

\begin{table*}[t]
\centering
\begin{tabular}{@{}lccccc@{}}
\hline
\bf User & 
\begin{tabular}[c]{@{}c@{}}\bf Whisper-\\\bf Small\end{tabular} & 
\begin{tabular}[c]{@{}c@{}}\bf Self Model\end{tabular} & 
\begin{tabular}[c]{@{}c@{}}\bf Common \end{tabular} & 
\begin{tabular}[c]{@{}c@{}}\bf ICI Model\\\bf (Avg. WER)\end{tabular} & 
\begin{tabular}[c]{@{}c@{}}\bf ICI Model\\\bf (Best WER)\end{tabular} \\ 
\hline
F01 & 83.33 & 41.67 & 53.57 & 41.27 & \textbf{36.11} \\
F03 & 43.37 & 30.47 & 34.56 & 29.54 & \textbf{28.67} \\
F04 & 13.96 & \textbf{9.01} & 12.55 & 10.10 & \textbf{9.01} \\ 
M01 & 99.47 & 43.32 & 65.39 & 44.31 & \textbf{37.43} \\ 
M02 & 81.67 & 37.78 & 66.82 & 36.03 & \textbf{35.00} \\
M03 & 7.08 & \textbf{6.60} & 9.77 & 9.10 & \textbf{6.60} \\ 
M04 & 149.34 & 59.21 & 83.83 & 57.05 & \textbf{50.66} \\
M05 & 89.29 & 64.29 & 67.86 & 64.80 & \textbf{57.14} \\ 
\hline
\textbf{Average} & 70.94 & 36.54 & 49.30 & 36.53 & \textbf{32.58} \\
\hline
\end{tabular}
\caption{Performance comparison across Encoder finetuned models (WER \%). 
Lower is better
The ICI Model (Avg. WER) column shows the average WER from training the user with every other normative model that excludes the user.
The last column chooses the best dysarthric idiosyncratic model for a user.}
\label{model-performance-table}
\end{table*}
 
As shown in \autoref{model-performance-table}, the Idiosyncratic models have an average WER of 36.54\%, whereas the Dysarthric Idiosyncratic models achieve a similar average WER of 36.53\% but a better best WER of 32.58\%. 
This improvement is observed for almost all users, suggesting that the speech patterns learned by the Dysarthric Normative model are effectively transferable to individual users during personalization.
Fine-tuning an idiosyncratic model from a Dysarthric Normative model yields better performance than starting directly from a Normative model.

From \autoref{model-performance-table} we found that the Dysarthric Idiosyncratic models improve WER by 54.07\% (70.94 → 32.58) compared to the Normative model. 
This gain is attributed to two factors: 30.5\% (70.94 → 49.30) of the improvement comes from learning common dysarthric speech patterns during the normative stage, while 23.57\% (49.30 → 32.58) is due to further adaptation to personalized speech patterns. 
This approach proves more effective than relying solely on personalized adaptation, as seen in the Idiosyncratic model, where the total improvement was only 48.49\% (70.94 → 36.54), derived entirely from personalized speech patterns.

Next, we examine \autoref{architecture-performance-table}. 
The LM-only fine-tuned Dysarthric Idiosyncratic models were initialized from Speech-only Dysarthric Normative models. 
However, their performance gains are minimal and significantly worse than Idiosyncratic models using encoder or full fine-tuning.

For full fine-tuning, the Dysarthric Idiosyncratic models were initialized from their Dysarthric Normative counterparts that also underwent full fine-tuning. 
Although these models performed better than the LM-only models, they still underperformed compared to the Speech-only models. 
This performance degradation can be attributed to the cascading effect caused by iterative LM tuning—first in the Dysarthric Normative model and then in the Dysarthric Idiosyncratic model—potentially leading to overfitting or instability in language modeling.

From these results, it is evident that the Speech Encoder plays the most crucial role in improving the normative models. 
For this dataset, full fine-tuning and encoder-only fine-tuning yield comparable results for Idiosyncratic models, making it difficult to draw definitive conclusions about their relative effectiveness. 
Further investigation with larger datasets may be necessary to determine whether one approach consistently outperforms the other.

\subsection{Effect of model size}

\autoref{architecture-performance-table-medium} shows our final results when we replace whisper-small models with whisper-medium.
The larger model shows a similar pattern where tuning the encoder yields better model performance.
The WERs improve for the larger model as is expected due to the larger learning capacity of the encoder.
The Idiosyncratic models for Whisper-Medium outperform the Dysarthric Idiosyncratic models for Whisper-Small.

On similar lines as the whisper-small model, for the whisper-medium models in \autoref{architecture-performance-table-medium}, we found that the Dysarthric Idiosyncratic models improve WER by 53.73\% (61.38 → 28.40) compared to the Normative model. 
This gain is attributed to two factors: 25.85\% (61.38 → 45.51) of the improvement comes from learning common dysarthric speech patterns during the normative stage, while 27.88\% (45.51 → 28.40) is due to further adaptation to personalized speech patterns. 
This approach proves more effective than relying solely on personalized adaptation, as seen in the Idiosyncratic model, where the total improvement was only 49.26\% (61.38 → 31.14), derived entirely from personalized speech patterns.

\subsection{Effect of train data size}

Since we have established that using a Speech-tuned Dysarthric Normative model is beneficial for training personalized models, the next experiment aimed to determine how much data is required for effective personalization when starting from a Dysarthric Normative model compared to training directly from scratch.

\begin{figure}[t]
\begin{center}
\includegraphics[width=\linewidth]{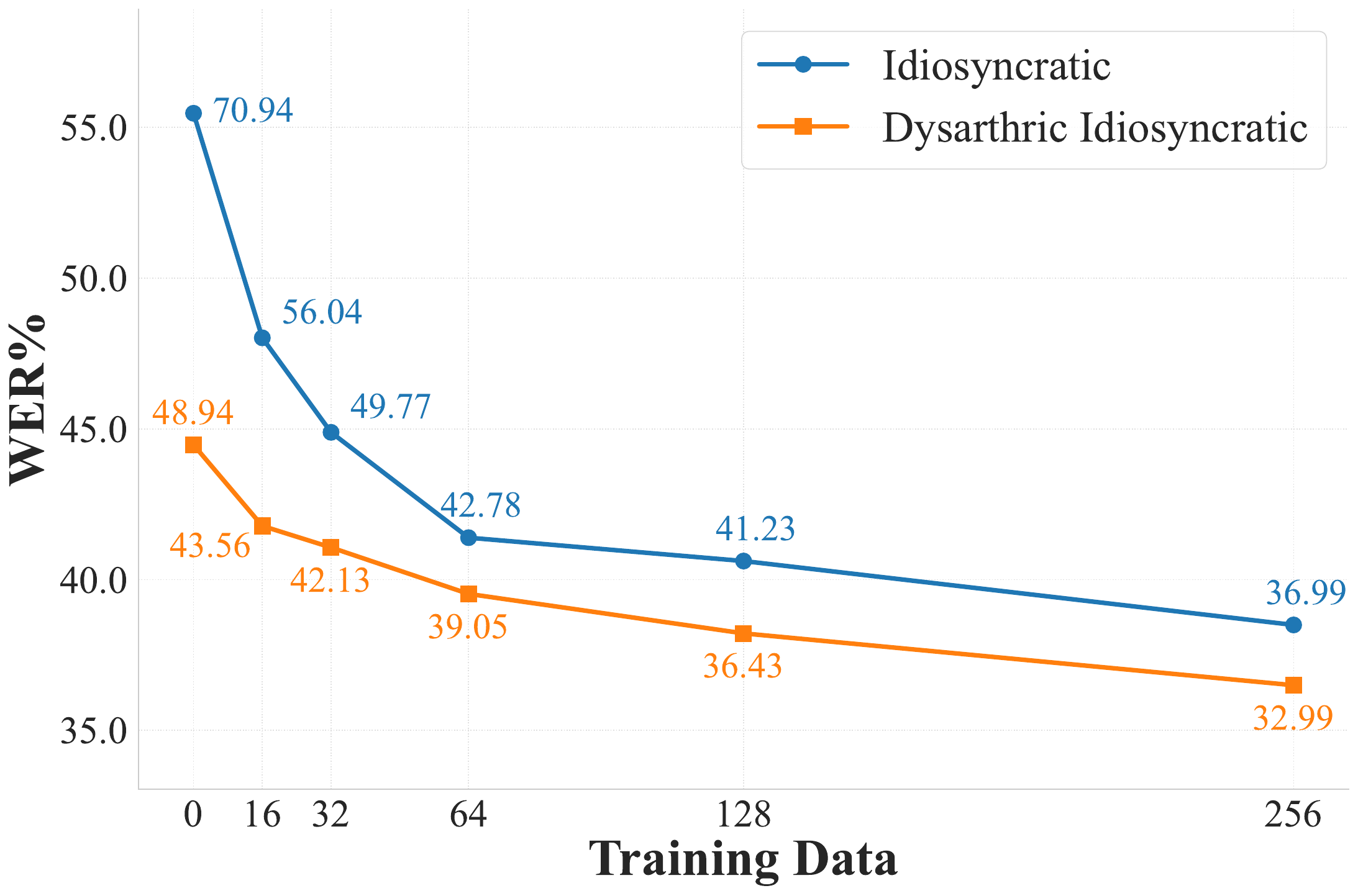}
\end{center}
\caption{Error as a function of training size for both idiosyncratic and dysarthric idiosyncratic. The benefits of the dysarthric idiosyncratic, which generalizes across dysarthric speakers, are larger at the smaller training set sizes but a benefit remains even with greater training sizes. }
\label{fig:performance-chart}
\end{figure}

Figure~\ref{fig:performance-chart} illustrates the relationship between training data size (X-axis) and average WER (Y-axis). 
Whisper-small was trained for each user using 16, 32, 64, 128, and 256 recordings. 
If a user had fewer than the specified number of recordings, all available training samples were used. 
The WER was computed using the full test dataset of each user at every step and then averaged.

The results indicate that when training a Dysarthric Idiosyncratic model, using only 128 recordings ($\sim$50\% of the full dataset) achieves better performance than training a personalized model with all 256 recordings from scratch. 
This finding suggests that by leveraging a Dysarthric Normative model, users can obtain a highly personalized ASR model with less than half the usual data collection effort, significantly reducing the burden of dataset creation while still achieving optimal recognition performance.


  
\subsection{Correlation of Model WERs and FDA scores}


\begin{figure}[t]
\begin{center}
\includegraphics[width=1.05\linewidth]{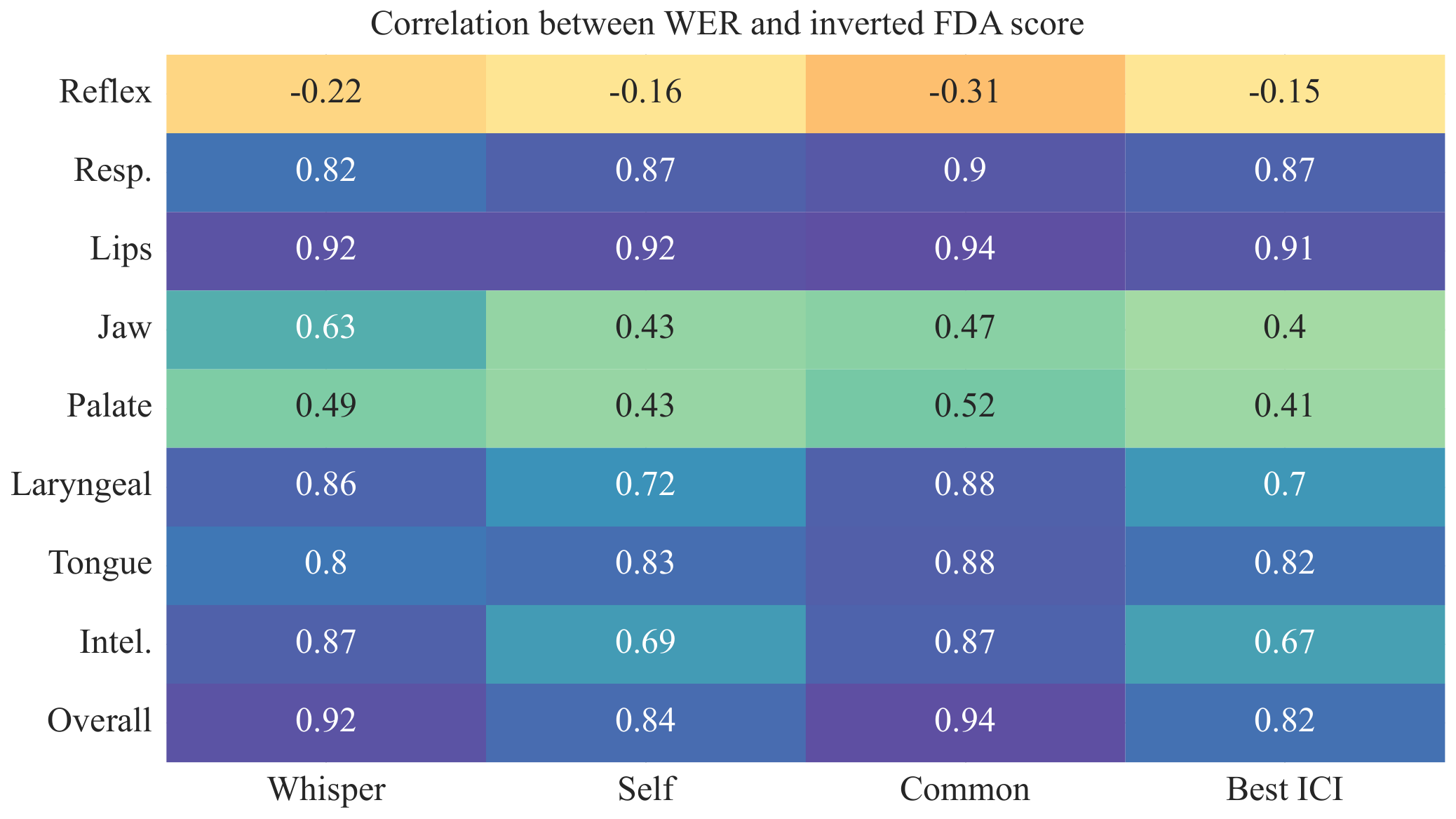}
\end{center}
\caption{Correlation between WER and inverted FDA score. 
We invert the FDA score to have a high score correspond to higher severity of dysarthria.
A higher value in a row means that the degree of imparity can explain the WER of the model more. 
A model that had learned dysarthric patterns would have a smaller correlation with inverted FDA scores, indicating that severity doesn't explain the model errors as much. 
}
\label{fig:corr_heatmap}
\end{figure}

While Word Error Rate (WER) provides a standard benchmark for transcription accuracy, it does not capture whether model errors are systematically related to the underlying motor-speech impairments of dysarthria. 
To address this, we examine correlations between model WERs and clinical Frenchay Dysarthria Assessment (FDA) scores. 
From a psychological measurement perspective, this serves as an evaluation of external validity: if model errors increase with clinical severity, then the model is sensitive to dysarthric impairment, whereas weaker correlations suggest that errors stem from other sources (e.g., model limitations unrelated to motor speech).

We find a high correlation between WER and inverted FDA scores\footnote{high score now correspond to high severity} in baseline models, indicating that transcription errors rise as speech quality worsens and are strongly tied to dysarthria-related disfluencies. 
However, when training on Idiosyncratic (Self) and Dysarthric Idiosyncratic (ICI) models, this correlation reduced. 
This suggests that the relationship between WER and speech disability weakens in these models —- implying that disability becomes less of a factor in transcription efficiency, which is a desired outcome. 
Except for Respiration and Tongue, all rows show a similar trend of improvement from Normative to Idiosyncratic models.

Taken together, this analysis complements standard WER results by offering a clinically grounded perspective: models with lower correlations to inverted FDA scores are not only more accurate, but also less constrained by the speaker’s impairment profile, highlighting their potential utility for individuals with dysarthria.

\section{Conclusions}\label{sec:conclusions}
Our study demonstrates that personalized fine-tuning remains critical for recognizing dysarthric speech, but can be made more efficient by leveraging dysarthric-normative pretraining and selectively adapting the speech encoder. 
By identifying the role of parameter subspaces in ASR models —- specifically the greater impact of tuning the speech encoder over the language decoder -- we enable a dysarthric-idiosyncratic approach to perform on par with, or better than, the widely used idiosyncratic models. 
To the best of our knowledge, this is the first work to show how combined modeling can outperform purely personalized strategies for disordered speech recognition under constrained data settings.

\section*{Limitations}



This study has several important limitations. 
First, our models were evaluated on a small number of speakers from the TORGO dataset. 
We selected TORGO because it uniquely reflects ecological validity through open-ended speech, unlike other dysarthric datasets that rely on scripted prompts. 
While this choice provides greater qualitative diversity, it also limits scalability and statistical power. 
Although we applied standard safeguards to reduce overfitting, our findings highlight the pressing need for larger, more representative datasets that capture naturalistic variation in dysarthric speech across populations and conditions. 
We view this work as a case study and a foundation for future large-scale validation.

Second, speaker-specific factors such as regional accent, dialect, or broader linguistic background were not controlled, despite their likely influence on transcription performance. 
Similarly, the limited dataset constrains the strength of our normative models; access to more diverse normative speech data would improve their robustness and comparability.

Finally, we did not examine incremental or longitudinal training strategies. 
Such approaches would be valuable for modeling the progressive trajectories of degenerative speech disorders, and may better reflect real-world use cases where systems adapt alongside an individual’s changing speech profile.

\section*{Ethical Considerations}

This work uses publicly available datasets containing dysarthric speech. 
All data used in this study were collected and released by their original creators and made available for research purposes. 
We ensured compliance with the dataset licenses and terms of use.

We acknowledge that dysarthric speech originates from individuals with medical conditions, and thus represents sensitive data.
While no personally identifiable information (PII) is present in the data, we took care to treat the speech recordings and associated metadata respectfully and strictly for the intended research purpose.

Our models are not designed for diagnostic use, and we caution against misuse of automatic systems for clinical decision-making without expert oversight.
Additionally, while this study focuses on improving transcription accuracy, we recognize the importance of inclusive AI development that does not reinforce biases against people with disabilities. 
Future work should prioritize user-centered evaluation and collaboration with affected communities.
Analysis of model behaviors~\cite{ganesan2024explaining} under varied settings can be a useful way to understand and explain the capabilities of these models.

\section*{Acknowledgments}
This study was funded by The National Institutes of Health, Smart and Connected Health, Grant NIH/NIMH R01 MH125702, and part funded by U01OH012476.
The conclusions contained herein are those of the authors and should not be interpreted as necessarily representing the official policies, either expressed or implied, of NSF, NIH, any other government organization, or the U.S. Government.


\bibliography{custom}

\begin{thebibliography}{45}
\providecommand{\natexlab}[1]{#1}

\bibitem[{A. et~al.(2016)A., T., and P.}]{tamil2016}
Mariya Celin~T. A., Nagarajan T., and Vijayalakshmi P. 2016.
\newblock \href {https://doi.org/10.1109/TENCON.2016.7848510} {{Dysarthric speech corpus in Tamil for rehabilitation research}}.
\newblock In \emph{2016 IEEE Region 10 Conference (TENCON)}, pages 2610--2613.

\bibitem[{Aw et~al.(2024)Aw, Montariol, AlKhamissi, Schrimpf, and Bosselut}]{aw2024instructiontuning}
Khai~Loong Aw, Syrielle Montariol, Badr AlKhamissi, Martin Schrimpf, and Antoine Bosselut. 2024.
\newblock \href {https://openreview.net/forum?id=nXNN0x4wbl} {Instruction-tuning aligns {LLM}s to the human brain}.
\newblock In \emph{First Conference on Language Modeling}.

\bibitem[{Baevski et~al.(2020)Baevski, Zhou, Mohamed, and Auli}]{baevski2020wav2vec}
Alexei Baevski, Henry Zhou, Abdelrahman Mohamed, and Michael Auli. 2020.
\newblock \href {https://arxiv.org/abs/2006.11477} {{wav2vec 2.0: A Framework for Self-Supervised Learning of Speech Representations}}.
\newblock \emph{CoRR}, abs/2006.11477.

\bibitem[{Baskar et~al.(2022)Baskar, Herzig, Nguyen, Diez, Polzehl, Burget, and Černocký}]{baskar2022speaker}
Murali~Karthick Baskar, Tim Herzig, Diana Nguyen, Mireia Diez, Tim Polzehl, Lukas Burget, and Jan Černocký. 2022.
\newblock \href {https://doi.org/10.21437/Interspeech.2022-10896} {{Speaker adaptation for Wav2vec2 based dysarthric ASR}}.
\newblock In \emph{Interspeech 2022}, pages 3403--3407.

\bibitem[{Bhat and Strik(2025)}]{bhat2025speech}
Chitralekha Bhat and Helmer Strik. 2025.
\newblock \href {https://doi.org/10.1044/2024\_JSLHR-23-00740} {{Speech Technology for Automatic Recognition and Assessment of Dysarthric Speech: An Overview}}.
\newblock \emph{Journal of Speech, Language, and Hearing Research}, 68(2):547--577.

\bibitem[{Doshi et~al.(2021)Doshi, Chen, Jiang, Zhang, Biadsy, Ramabhadran, Chu, Rosenberg, and Moreno}]{doshi2021extending}
Rohan Doshi, Youzheng Chen, Liyang Jiang, Xia Zhang, Fadi Biadsy, Bhuvana Ramabhadran, Fang Chu, Andrew Rosenberg, and Pedro~J. Moreno. 2021.
\newblock \href {https://doi.org/10.1109/ICASSP39728.2021.9414644} {{Extending Parrotron: An End-to-End, Speech Conversion and Speech Recognition Model for Atypical Speech}}.
\newblock In \emph{ICASSP 2021 - 2021 IEEE International Conference on Acoustics, Speech and Signal Processing (ICASSP)}, pages 6988--6992.

\bibitem[{Enderby(2011)}]{enderby2011}
Pamela Enderby. 2011.
\newblock \href {https://doi.org/10.3109/13682828009112541} {{The Frenchay Dysarthria Assessment}}.
\newblock \emph{International Journal of Language \& Communication Disorders}, 15:165 -- 173.

\bibitem[{Goldstein et~al.(2025)Goldstein, Wang, Niekerken, Schain, Zada, Aubrey, Sheffer, Nastase, Gazula, Singh, Rao, Choe, Kim, Doyle, Friedman, Devore, Dugan, Hassidim, Brenner, and Hasson}]{goldstein2025unified}
Ariel Goldstein, Haocheng Wang, Leonard Niekerken, Mariano Schain, Zaid Zada, Bobbi Aubrey, Tom Sheffer, Samuel Nastase, Harshvardhan Gazula, Aditi Singh, Aditi Rao, Gina Choe, Catherine Kim, Werner Doyle, Daniel Friedman, Sasha Devore, Patricia Dugan, Avinatan Hassidim, Michael Brenner, and Uri Hasson. 2025.
\newblock \href {https://doi.org/10.1038/s41562-025-02105-9} {A unified acoustic-to-speech-to-language embedding space captures the neural basis of natural language processing in everyday conversations}.
\newblock \emph{Nature Human Behaviour}, 9:1041--1055.

\bibitem[{Green et~al.(2021)Green, MacDonald, Jiang, Cattiau, Heywood, Cave, Seaver, Ladewig, Tobin, Brenner, Nelson, and Tomanek}]{green2021automatic}
Jordan~R. Green, Robert~L. MacDonald, Pan-Pan Jiang, Julie Cattiau, Rus Heywood, Richard Cave, Katie Seaver, Marilyn~A. Ladewig, Jimmy Tobin, Michael~P. Brenner, Philip~C. Nelson, and Katrin Tomanek. 2021.
\newblock \href {https://doi.org/10.21437/Interspeech.2021-1384} {{Automatic Speech Recognition of Disordered Speech: Personalized Models Outperforming Human Listeners on Short Phrases}}.
\newblock In \emph{Interspeech 2021}, pages 4778--4782.

\bibitem[{Hermann and Magimai-Doss(2023)}]{hermann2023few}
Enno Hermann and Mathew Magimai-Doss. 2023.
\newblock \href {https://doi.org/10.21437/Interspeech.2023-2481} {Few-shot dysarthric speech recognition with text-to-speech data augmentation}.
\newblock In \emph{Interspeech 2023}, pages 156--160.

\bibitem[{Hoffmann et~al.(2022)Hoffmann, Borgeaud, Mensch, Buchatskaya, Cai, Rutherford, de~Las~Casas, Hendricks, Welbl, Clark, Hennigan, Noland, Millican, van~den Driessche, Damoc, Guy, Osindero, Simonyan, Elsen, Rae, Vinyals, and Sifre}]{hoffmann-etal-2022}
Jordan Hoffmann, Sebastian Borgeaud, Arthur Mensch, Elena Buchatskaya, Trevor Cai, Eliza Rutherford, Diego de~Las~Casas, Lisa~Anne Hendricks, Johannes Welbl, Aidan Clark, Tom Hennigan, Eric Noland, Katie Millican, George van~den Driessche, Bogdan Damoc, Aurelia Guy, Simon Osindero, Karen Simonyan, Erich Elsen, and 3 others. 2022.
\newblock \href {https://arxiv.org/abs/2203.15556} {{Training Compute-Optimal Large Language Models}}.
\newblock \emph{Preprint}, arXiv:2203.15556.

\bibitem[{Hsieh et~al.(2024)Hsieh, Choi, and Kim}]{hsieh2024multimodal}
Tsun-An Hsieh, Heeyoul Choi, and Minje Kim. 2024.
\newblock \href {https://doi.org/10.21437/Interspeech.2024-444} {{Multimodal Representation Loss Between Timed Text and Audio for Regularized Speech Separation}}.
\newblock In \emph{Interspeech 2024}, pages 1300--1304.

\bibitem[{Hu et~al.(2022{\natexlab{a}})Hu, yelong shen, Wallis, Allen-Zhu, Li, Wang, Wang, and Chen}]{hu2022lora}
Edward~J Hu, yelong shen, Phillip Wallis, Zeyuan Allen-Zhu, Yuanzhi Li, Shean Wang, Lu~Wang, and Weizhu Chen. 2022{\natexlab{a}}.
\newblock \href {https://openreview.net/forum?id=nZeVKeeFYf9} {Lo{RA}: Low-rank adaptation of large language models}.
\newblock In \emph{International Conference on Learning Representations}.

\bibitem[{Hu et~al.(2022{\natexlab{b}})Hu, Xie, Cui, Deng, Liu, Yu, Geng, Liu, and Meng}]{hu2022neural}
Shoukang Hu, Xurong Xie, Mingyu Cui, Jiajun Deng, Shansong Liu, Jianwei Yu, Mengzhe Geng, Xunying Liu, and Helen Meng. 2022{\natexlab{b}}.
\newblock \href {https://doi.org/10.1109/TASLP.2022.3153253} {{Neural Architecture Search for LF-MMI Trained Time Delay Neural Networks}}.
\newblock \emph{IEEE/ACM Transactions on Audio, Speech, and Language Processing}, 30:1093--1107.

\bibitem[{Hu et~al.(2024)Hu, Xie, Geng, Jin, Deng, Li, Wang, Cui, Wang, Meng, and Liu}]{hu2024self}
Shujie Hu, Xurong Xie, Mengzhe Geng, Zengrui Jin, Jiajun Deng, Guinan Li, Yi~Wang, Mingyu Cui, Tianzi Wang, Helen Meng, and Xunying Liu. 2024.
\newblock \href {https://doi.org/10.1109/TASLP.2024.3422839} {{Self-Supervised ASR Models and Features for Dysarthric and Elderly Speech Recognition}}.
\newblock \emph{IEEE/ACM Transactions on Audio, Speech, and Language Processing}, 32:3561--3575.

\bibitem[{Javanmardi et~al.(2024)Javanmardi, Kadiri, and Alku}]{javanmardi2024exploring}
Farhad Javanmardi, Sudarsana~Reddy Kadiri, and Paavo Alku. 2024.
\newblock \href {https://doi.org/10.1109/JBHI.2024.3392829} {{Exploring the Impact of Fine-Tuning the Wav2vec2 Model in Database-Independent Detection of Dysarthric Speech}}.
\newblock \emph{IEEE Journal of Biomedical and Health Informatics}, 28(8):4951--4962.

\bibitem[{Javanmardi et~al.(2023)Javanmardi, Tirronen, Kodali, Kadiri, and Alku}]{javanmardi2023wav}
Farhad Javanmardi, Saska Tirronen, Manila Kodali, Sudarsana~Reddy Kadiri, and Paavo Alku. 2023.
\newblock \href {https://doi.org/10.1109/ICASSP49357.2023.10094857} {{Wav2vec-Based Detection and Severity Level Classification of Dysarthria From Speech}}.
\newblock In \emph{ICASSP 2023 - 2023 IEEE International Conference on Acoustics, Speech and Signal Processing (ICASSP)}, pages 1--5.

\bibitem[{Kaplan et~al.(2020)Kaplan, McCandlish, Henighan, Brown, Chess, Child, Gray, Radford, Wu, and Amodei}]{kaplan2020scaling}
Jared Kaplan, Sam McCandlish, Tom Henighan, Tom~B. Brown, Benjamin Chess, Rewon Child, Scott Gray, Alec Radford, Jeffrey Wu, and Dario Amodei. 2020.
\newblock \href {https://arxiv.org/abs/2001.08361} {{Scaling Laws for Neural Language Models}}.
\newblock \emph{Preprint}, arXiv:2001.08361.

\bibitem[{Kim et~al.(2008)Kim, Hasegawa-Johnson, Perlman, Gunderson, Huang, Watkin, and Frame}]{kim2008uaspeech}
Heejin Kim, Mark Hasegawa-Johnson, Adrienne Perlman, Jon Gunderson, Thomas~S. Huang, Kenneth Watkin, and Simone Frame. 2008.
\newblock \href {https://doi.org/10.21437/Interspeech.2008-480} {{Dysarthric speech database for universal access research}}.
\newblock In \emph{Interspeech 2008}, pages 1741--1744.

\bibitem[{M\"{u}ller-Eberstein et~al.(2024)M\"{u}ller-Eberstein, Yee, Yang, Mantena, and Lea}]{muller-eberstein2024}
Max M\"{u}ller-Eberstein, Dianna Yee, Karren Yang, Gautam~Varma Mantena, and Colin Lea. 2024.
\newblock \href {https://doi.org/10.1162/tacl_a_00696} {{Hypernetworks for Personalizing ASR to Atypical Speech}}.
\newblock \emph{Transactions of the Association for Computational Linguistics}, 12:1182--1196.

\bibitem[{Nguyen et~al.(2024)Nguyen, Fredouille, Ghio, Balaguer, and Woisard}]{nguyen2024exploring}
Tuan Nguyen, Corinne Fredouille, Alain Ghio, Mathieu Balaguer, and Virginie Woisard. 2024.
\newblock \href {https://doi.org/10.1109/SLT61566.2024.10832202} {{Exploring ASR-Based WAV2VEC2 for Automated Speech Disorder Assessment: Insights and Analysis}}.
\newblock In \emph{2024 IEEE Spoken Language Technology Workshop (SLT)}, pages 975--982.

\bibitem[{Qi and {Van hamme}(2023)}]{qi2023interspeech}
Jinzi Qi and Hugo {Van hamme}. 2023.
\newblock \href {https://doi.org/10.21437/Interspeech.2023-1627} {{Parameter-efficient Dysarthric Speech Recognition Using Adapter Fusion and Householder Transformation}}.
\newblock In \emph{Interspeech 2023}, pages 151--155.

\bibitem[{Qi and Van~hamme(2025)}]{qi2025model}
Jinzi Qi and Hugo Van~hamme. 2025.
\newblock \href {https://doi.org/10.3390/app15042006} {{A Study on Model Training Strategies for Speaker-Independent and Vocabulary-Mismatched Dysarthric Speech Recognition}}.
\newblock \emph{Applied Sciences}, 15(4).

\bibitem[{Radford et~al.(2022)Radford, Kim, Xu, Brockman, McLeavey, and Sutskever}]{radford2022robustspeechrecognitionlargescale}
Alec Radford, Jong~Wook Kim, Tao Xu, Greg Brockman, Christine McLeavey, and Ilya Sutskever. 2022.
\newblock \href {https://arxiv.org/abs/2212.04356} {{Robust Speech Recognition via Large-Scale Weak Supervision}}.
\newblock \emph{Preprint}, arXiv:2212.04356.

\bibitem[{Rudzicz et~al.(2010)Rudzicz, Namasivayam, and Wolff}]{rudzicz2012torgo}
Frank Rudzicz, Aravind Namasivayam, and Talya Wolff. 2010.
\newblock \href {https://doi.org/10.1007/s10579-011-9145-0} {{The TORGO database of acoustic and articulatory speech from speakers with dysarthria}}.
\newblock \emph{Language Resources and Evaluation}, 46:1--19.

\bibitem[{Shor et~al.(2019)Shor, Emanuel, Lang, Tuval, Brenner, Cattiau, Vieira, McNally, Charbonneau, Nollstadt, Hassidim, and Matias}]{shor2019personalizing}
Joel Shor, Dotan Emanuel, Oran Lang, Omry Tuval, Michael Brenner, Julie Cattiau, Fernando Vieira, Maeve McNally, Taylor Charbonneau, Melissa Nollstadt, Avinatan Hassidim, and Yossi Matias. 2019.
\newblock \href {https://doi.org/10.21437/Interspeech.2019-1427} {{Personalizing ASR for Dysarthric and Accented Speech with Limited Data}}.
\newblock In \emph{Interspeech 2019}, pages 784--788.

\bibitem[{Takashima et~al.(2024{\natexlab{a}})Takashima, Otani, Aihara, Takiguchi, and Taguchi}]{takashima2024self}
Ryoichi Takashima, Takeru Otani, Ryo Aihara, Tetsuya Takiguchi, and Shinya Taguchi. 2024{\natexlab{a}}.
\newblock \href {https://doi.org/10.1145/3663548.3688536} {Self-supervised learning using unlabeled speech with multiple types of speech disorder for disordered speech recognition}.
\newblock In \emph{Proceedings of the 26th International ACM SIGACCESS Conference on Computers and Accessibility}, ASSETS '24, New York, NY, USA. Association for Computing Machinery.

\bibitem[{Takashima et~al.(2024{\natexlab{b}})Takashima, Sawa, Aihara, Takiguchi, and Imai}]{takashima2024dysarthric}
Ryoichi Takashima, Yuya Sawa, Ryo Aihara, Tetsuya Takiguchi, and Yoshie Imai. 2024{\natexlab{b}}.
\newblock \href {https://doi.org/10.1109/ACCESS.2024.3374874} {{Dysarthric Speech Recognition Using Pseudo-Labeling, Self-Supervised Feature Learning, and a Joint Multi-Task Learning Approach}}.
\newblock \emph{IEEE Access}, 12:36990--36999.

\bibitem[{Takashima et~al.(2020{\natexlab{a}})Takashima, Takiguchi, and Ariki}]{takashima2020two}
Ryoichi Takashima, Tetsuya Takiguchi, and Yasuo Ariki. 2020{\natexlab{a}}.
\newblock \href {https://doi.org/10.1109/ICASSP40776.2020.9053725} {{Two-Step Acoustic Model Adaptation for Dysarthric Speech Recognition}}.
\newblock In \emph{ICASSP 2020 - 2020 IEEE International Conference on Acoustics, Speech and Signal Processing (ICASSP)}, pages 6104--6108.

\bibitem[{Takashima et~al.(2020{\natexlab{b}})Takashima, Takashima, Takiguchi, and Ariki}]{takashima2020dysarthric}
Yuki Takashima, Ryoichi Takashima, Tetsuya Takiguchi, and Yasuo Ariki. 2020{\natexlab{b}}.
\newblock \href {https://doi.org/10.21437/Interspeech.2020-2267} {{Dysarthric Speech Recognition Based on Deep Metric Learning}}.
\newblock In \emph{Interspeech 2020}, pages 4796--4800.

\bibitem[{Tomanek et~al.(2021{\natexlab{a}})Tomanek, Beaufays, Cattiau, Chandorkar, and Sim}]{tomanek2021on}
Katrin Tomanek, Françoise Beaufays, Julie Cattiau, Angad Chandorkar, and Khe~Chai Sim. 2021{\natexlab{a}}.
\newblock \href {https://arxiv.org/abs/2106.10259} {{On-Device Personalization of Automatic Speech Recognition Models for Disordered Speech}}.
\newblock \emph{Preprint}, arXiv:2106.10259.

\bibitem[{Tomanek et~al.(2023)Tomanek, Seaver, Jiang, Cave, Harrell, and Green}]{tomanek2023analysis}
Katrin Tomanek, Katie Seaver, Pan-Pan Jiang, Richard Cave, Lauren Harrell, and Jordan Green. 2023.
\newblock \href {https://doi.org/10.1109/ICASSP49357.2023.10097195} {{An Analysis of Degenerating Speech Due to Progressive Dysarthria on ASR Performance}}.
\newblock In \emph{ICASSP 2023 - 2023 IEEE International Conference on Acoustics, Speech and Signal Processing (ICASSP)}, pages 1--5.

\bibitem[{Tomanek et~al.(2021{\natexlab{b}})Tomanek, Zayats, Padfield, Vaillancourt, and Biadsy}]{tomanek2021residual}
Katrin Tomanek, Vicky Zayats, Dirk Padfield, Kara Vaillancourt, and Fadi Biadsy. 2021{\natexlab{b}}.
\newblock \href {https://arxiv.org/abs/2109.06952} {{Residual Adapters for Parameter-Efficient ASR Adaptation to Atypical and Accented Speech}}.
\newblock \emph{Preprint}, arXiv:2109.06952.

\bibitem[{Tuckute et~al.(2024)Tuckute, Sathe, Srikant, Taliaferro, Wang, Schrimpf, Kay, and Fedorenko}]{tuckute2024driving}
Greta Tuckute, Aalok Sathe, Shashank Srikant, Maya Taliaferro, Mingye Wang, Martin Schrimpf, Kendrick Kay, and Evelina Fedorenko. 2024.
\newblock \href {https://doi.org/10.1038/s41562-023-01783-7} {Driving and suppressing the human language network using large language models}.
\newblock \emph{Nature Human Behaviour}, 8:1--18.

\bibitem[{Turrisi et~al.(2021)Turrisi, Braccia, Emanuele, Giulietti, Pugliatti, Sensi, and andLeonardo Badino}]{turrisi2021easycall}
Rosanna Turrisi, Arianna Braccia, Marco Emanuele, Simone Giulietti, Maura Pugliatti, Mariachiara Sensi, and Luciano~Fadiga andLeonardo Badino. 2021.
\newblock \href {https://doi.org/10.21437/Interspeech.2021-549} {{EasyCall corpus: a dysarthric speech dataset}}.
\newblock In \emph{Interspeech 2021}, pages 41--45.

\bibitem[{V~Ganesan et~al.(2021)V~Ganesan, Matero, Ravula, Vu, and Schwartz}]{v-ganesan-etal-2021-empirical}
Adithya V~Ganesan, Matthew Matero, Aravind~Reddy Ravula, Huy Vu, and H.~Andrew Schwartz. 2021.
\newblock \href {https://doi.org/10.18653/v1/2021.naacl-main.357} {Empirical evaluation of pre-trained transformers for human-level {NLP}: The role of sample size and dimensionality}.
\newblock In \emph{Proceedings of the 2021 Conference of the North American Chapter of the Association for Computational Linguistics: Human Language Technologies}, pages 4515--4532, Online. Association for Computational Linguistics.

\bibitem[{V~Ganesan et~al.(2024)V~Ganesan, Varadarajan, Lal, Eijsbroek, Kjell, Kjell, Dhanasekaran, Stade, Eichstaedt, Boyd et~al.}]{ganesan2024explaining}
Adithya V~Ganesan, Vasudha Varadarajan, Yash~Kumar Lal, Veerle~C Eijsbroek, Katarina Kjell, Oscar~NE Kjell, Tanuja Dhanasekaran, Elizabeth~C Stade, Johannes~C Eichstaedt, Ryan~L Boyd, and 1 others. 2024.
\newblock Explaining gpt-4's schema of depression using machine behavior analysis.
\newblock \emph{arXiv preprint arXiv:2411.13800}.

\bibitem[{Vachhani et~al.(2017)Vachhani, Bhat, Das, and Kopparapu}]{vachhani2017deep}
Bhavik Vachhani, Chitralekha Bhat, Biswajit Das, and Sunil~Kumar Kopparapu. 2017.
\newblock \href {https://doi.org/10.21437/Interspeech.2017-1318} {{Deep Autoencoder Based Speech Features for Improved Dysarthric Speech Recognition}}.
\newblock In \emph{Interspeech 2017}, pages 1854--1858.

\bibitem[{Violeta et~al.(2022)Violeta, Huang, and Toda}]{violeta2022investigating}
Lester~Phillip Violeta, Wen~Chin Huang, and Tomoki Toda. 2022.
\newblock \href {https://doi.org/10.21437/Interspeech.2022-10043} {{Investigating Self-supervised Pretraining Frameworks for Pathological Speech Recognition}}.
\newblock In \emph{Interspeech 2022}, pages 41--45.

\bibitem[{Wan et~al.(2024)Wan, Sun, Kang, Li, Guo, Gao, and Wang}]{cdsd2023}
Yan Wan, Mengyi Sun, Xinchen Kang, Jingting Li, Pengfei Guo, Ming Gao, and Su-Jing Wang. 2024.
\newblock \href {https://doi.org/10.21437/interspeech.2024-1597} {{CDSD: Chinese Dysarthria Speech Database}}.
\newblock In \emph{Interspeech 2024}, page 4109–4113. ISCA.

\bibitem[{Wang et~al.(2021)Wang, Yu, Wu, Sun, Liu, and Meng}]{wang2021improved}
Disong Wang, Jianwei Yu, Xixin Wu, Lifa Sun, Xunying Liu, and Helen Meng. 2021.
\newblock \href {https://doi.org/10.1109/ISCSLP49672.2021.9362068} {{Improved End-to-End Dysarthric Speech Recognition via Meta-learning Based Model Re-initialization}}.
\newblock In \emph{2021 12th International Symposium on Chinese Spoken Language Processing (ISCSLP)}, pages 1--5.

\bibitem[{Wolf et~al.(2020)Wolf, Debut, Sanh, Chaumond, Delangue, Moi, Cistac, Rault, Louf, Funtowicz, Davison, Shleifer, von Platen, Ma, Jernite, Plu, Xu, Le~Scao, Gugger, Drame, Lhoest, and Rush}]{wolf-etal-2020-transformers}
Thomas Wolf, Lysandre Debut, Victor Sanh, Julien Chaumond, Clement Delangue, Anthony Moi, Pierric Cistac, Tim Rault, Remi Louf, Morgan Funtowicz, Joe Davison, Sam Shleifer, Patrick von Platen, Clara Ma, Yacine Jernite, Julien Plu, Canwen Xu, Teven Le~Scao, Sylvain Gugger, and 3 others. 2020.
\newblock \href {https://doi.org/10.18653/v1/2020.emnlp-demos.6} {{Transformers: State-of-the-Art Natural Language Processing}}.
\newblock In \emph{Proceedings of the 2020 Conference on Empirical Methods in Natural Language Processing: System Demonstrations}, pages 38--45, Online. Association for Computational Linguistics.

\bibitem[{Wu et~al.(2021)Wu, Zong, Sun, and Zhao}]{wu2021sequential}
Lidan Wu, Daoming Zong, Shiliang Sun, and Jing Zhao. 2021.
\newblock \href {https://doi.org/10.1109/ICASSP39728.2021.9415017} {{A Sequential Contrastive Learning Framework for Robust Dysarthric Speech Recognition}}.
\newblock In \emph{ICASSP 2021 - 2021 IEEE International Conference on Acoustics, Speech and Signal Processing (ICASSP)}, pages 7303--7307.

\bibitem[{Xiong et~al.(2019)Xiong, Barker, and Christensen}]{xiong2019phonetic}
Feifei Xiong, Jon Barker, and Heidi Christensen. 2019.
\newblock \href {https://doi.org/10.1109/ICASSP.2019.8683091} {{Phonetic Analysis of Dysarthric Speech Tempo and Applications to Robust Personalised Dysarthric Speech Recognition}}.
\newblock In \emph{ICASSP 2019 - 2019 IEEE International Conference on Acoustics, Speech and Signal Processing (ICASSP)}, pages 5836--5840.

\bibitem[{Yang et~al.(2025)Yang, Wu, Liu, Liu, Zhou, Wang, Wang, Su, Yan, and Wang}]{yang2025feature}
Yudong Yang, Xinyi Wu, Xiaokang Liu, Jiang Liu, Jingdong Zhou, Rennan Wang, Xin Wang, Rongfeng Su, Nan Yan, and Lan Wang. 2025.
\newblock \href {https://doi.org/10.1007/978-981-96-1151-5_28} {{Feature Extraction Method Based on Contrastive Learning for Dysarthria Detection}}.
\newblock In \emph{Int. Conf. Social Robotics}, pages 272--281. Springer, Singapore.

\end{thebibliography}

\appendix

\section*{Appendix}

\section{Dysarthric Normative Model: Leave-One-Out Cross-Validation}

The dysarthric normative model was built using a leave-one-out cross-validation approach across all eight speakers in the TORGO dataset. To make the process more transparent, we describe it below in pseudocode form.

        
        

\begin{figure*}[t]
\begin{lstlisting}[language=Python]
scores = []
for test_user in all_users:
    user_score = []
    for dev_user in all_users - {test_user}:
        train_users = all_users - {test_user, dev_user}
        train_data = data[train_users]
        dev_data = data[dev_user]
        test_data = data[test_user]

        model.fit(train=train_data, dev=dev_data)
        wer = model.eval(test_data)
        user_score.append(wer)

    scores.append(user_score)
\end{lstlisting}
\caption{Algorithm used for Leave-One-Out Cross-Validation of Dysarthric Normative models.}
\end{figure*}

This process results in a total of 56 models 
($8$ test users $\times$ $7$ dev users), 
ensuring speaker-independent evaluation and reducing overfitting.

\section*{Responsible NLP Research Checklist}

\subsection*{A. Limitations and Potential Risks}
\begin{itemize}
    \item \textbf{A1. Limitations Section:} Yes. This paper has a limitations section.
    \item \textbf{A2. Potential Risks:} Yes. Ethical Considerations.
\end{itemize}

\subsection*{B. Use or Creation of Scientific Artifacts}
\begin{itemize}
    \item \textbf{B. Use or Create Scientific Artifacts:} Yes.
    \item \textbf{B1. Cite Creators of Artifacts:} Yes. \S\ref{sec:dataset}, \S\ref{sec:experiments}.
    \item \textbf{B2. Discuss the License for Artifacts:} Yes. Ethical Consideration.
    \item \textbf{B3. Artifact Use Consistent with Intended Use:} Yes. Ethical Considerations.
    \item \textbf{B4. Data Contains Personally Identifying Info or Offensive Content:} No. Anonymized Data.
    \item \textbf{B5. Documentation of Artifacts:} Yes. \S\ref{sec:dataset}.
    \item \textbf{B6. Statistics for Data:} Yes. \S\ref{sec:dataset}.
\end{itemize}

\subsection*{C. Computational Experiments}
\begin{itemize}
    \item \textbf{C. Computational Experiments:} Yes.
    \item \textbf{C1. Model Size and Budget:} Yes. \S\ref{sec:experiments}.
    \item \textbf{C2. Experimental Setup and Hyperparameters:} Yes. \S \ref{sec:experiments}.
    \item \textbf{C3. Descriptive Statistics:} Yes. \S \ref{sec:results}.
    \item \textbf{C4. Parameters for Packages:} Yes. \S \ref{sec:experiments}.
\end{itemize}

\subsection*{D. Human Subjects Including Annotators}
\begin{itemize}
    \item \textbf{D. Human Subjects:} No.
    \item \textbf{D1. Instructions Given to Participants:} N/A.
    \item \textbf{D2. Recruitment and Payment:} N/A.
    \item \textbf{D3. Data Consent:} N/A.
    \item \textbf{D4. Ethics Review Board Approval:} N/A.
    \item \textbf{D5. Characteristics of Annotators:} N/A.
\end{itemize}

\subsection*{E. AI Assistants in Research or Writing}
\begin{itemize}
    \item \textbf{E. AI Assistants in Research or Writing:} Yes.
    \item \textbf{E1. Information About Use of AI Assistants:} No. Used as copilot and text correction tool.
\end{itemize}

\end{document}